\documentclass[preprint2]{aastex}
\usepackage{aastexug}

\newcommand{\deltavec}{\mbox{\boldmath $\delta$}}
\newcommand{\thetavec}{\mbox{\boldmath $\theta$}}
\newcommand{\uvec}{\mbox{\boldmath $u$}}
\newcommand{\xvec}{\mbox{\boldmath $x$}}
\newcommand{\yvec}{\mbox{\boldmath $y$}}

\def\plotfiddle#1#2#3#4#5#6#7{\centering \leavevmode
\vbox to#2{\rule{0pt}{#2}}
\includegraphics{#1}}

\begin{document}

\title{Astrometric Method to Break the Photometric Degeneracy between  
       Binary-source and Planetary Microlensing Perturbations}

\author{Cheongho Han}
\affil{Department of Physics,
Chungbuk National University, Chongju 361-763, Korea}
\email{cheongho@astroph.chungbuk.ac.kr}

\begin{abstract}
An extra-solar planet can be detected by microlensing because the planet 
can perturb the smooth lensing light curve created by the primary lens.  
However, it was shown by Gaudi that a subset of binary-source events can 
produce light curves that closely resemble those produced by a significant 
fraction of planet/star lens systems, causing serious contamination of 
a sample of suspected planetary systems detected via microlensing.  
In this paper, we show that if a lensing event is observed astrometrically, 
one can unambiguously break the photometric degeneracy between binary-source 
and planetary lensing perturbations.  This is possible because while the 
planet-induced perturbation in the trajectory of the lensed source image 
centroid shifts points away from the opening of the unperturbed elliptical 
trajectory, while the perturbation induced by the binary source companion 
points {\it always} towards the opening.  Therefore, astrometric microlensing 
observations by using future high-precision interferometers will be important
for solid confirmation of microlensing planet detections.
\end{abstract}

\section{Introduction}

Over the past decade, microlensing has developed into a powerful tool in 
various aspects of astrophysics besides its original use of searching for 
Galactic dark matter in the form of massive compact halo objects.  One 
important application is the detection of extra-solar planets \citep{mao91, 
gould92}.  A planet can be detected by microlensing because the planet 
can perturb the smooth lensing light curve created by the primary lens.  
Since typical planet-induced perturbations last only a few hours to days, 
detecting them requires intensive monitoring of events.  Currently, two 
groups \citep{rhie00, albrow01} are conducting follow-up observations of 
ongoing events alerted by the survey experiments \citep{alcock00, 
lasserre00, soszynski01, bond01} to increase the planet detection rate by 
improving the monitoring frequency and photometric accuracy.  Once the 
perturbation is detected, one can determine the mass ratio and the 
projected separation of the planet.

However, a mere detection of a short-lived perturbation in the lensing 
light curve does not guarantee the detection of a planet.  This is because 
similar anomalies can also occur due to totally different physical reasons 
\citep{gaudi97, dominik99}.  One well known type of events that produce 
light curves closely resembling those produced by more than half of 
planet/star lens systems are a subset of binary-source events where the 
source companion has a very low flux ratio to the primary source: 
planet/binary-source degeneracy \citep{gaudi98}.  For these events, the 
perturbation occurs when the lens passes closely to the source companion.
\citet{gaudi98} showed that the probability for this type of perturbations
can be comparable to the detection probability of Jupiter-mass planets.
Therefore, unless this degeneracy is broken and the true cause of the 
perturbation is determined, a sample of suspected planetary systems 
detected via microlensing will be seriously contaminated.

As a new method of detecting planets via microlensing, \citet{safizadeh99} 
proposed to conduct astrometric follow-up observations of lensing events 
by using several planned high-precision interferometers such as the {\it 
Space Interferometry Mission} [SIM, \citet{unwin97}] and those to be 
mounted on the Keck \citep{colavita98} and the VLT \citep{mariotti98}.  
From astrometric microlensing observations by using these interferometers, 
one can measure the displacements in the source star image center of 
light with respect to its unlensed position, $\deltavec$ \citep{miyamoto95, 
hog95}.  The trajectory of the centroid shifts (astrometric curve) caused 
by a single point-mass lens is an ellipse \citep{walker95, jeong99}.  A 
planet can be detected because it can perturb the elliptical astrometric 
curve of the single lens event, which is analogous to the perturbation 
in the light curve.  Adding astrometric information to the photometric 
light curve greatly helps in determining the planetary mass and projected 
separation, because the lensing behaviors of photometric and astrometric 
perturbations are strongly correlated \citep{safizadeh99, han01b}.

In this paper, we investigate whether one can break the planet/binary-source 
degeneracy in lensing light curves from astrometric microlensing observations.  
In \S\ 2, we briefly describe the basics of microlensing.  In \S\ 3, we 
show that the degeneracy can be unambiguously broken astrometrically by 
demonstrating the fundamental differences between the astrometric curves 
resulting from a photometrically degenerate case of binary-source and 
planetary lensing events.  We conclude in \S\ 5.

\section{Basics of Microlensing}
\subsection{Standard events}
If a single source located at $\thetavec_{\rm S}$ on the projected plane 
is lensed by a coplanar $N$ point-mass lens system, where the individual 
masses and locations are $m_j$ and $\thetavec_{{\rm L},j}$, the positions 
of the resulting images $\thetavec$ are obtained by solving the lens 
equation of the form
\begin{equation}
\thetavec_{\rm S}=\thetavec - {\theta_{\rm E}^2\over m}
          \sum_{j=1}^{N}m_{j} { \thetavec-\thetavec_{{\rm L},j}\over 
          {\left\vert \thetavec-\thetavec_{{\rm L},j}\right\vert}^2},
\end{equation}
where $m=\sum_{j=1}^N m_j$ is the total mass of the lens system and
$\theta_{\rm E}$ is the angular Einstein ring radius.  The Einstein
ring radius is related to the physical parameters of the lens system by
\begin{equation}
\theta_{\rm E} = \sqrt{4Gm\over c^2} 
\left({1\over D_{\rm ol}}-{1\over D_{\rm os}}\right)^{1/2},
\end{equation}
where $D_{\rm ol}$ and $D_{\rm os}$ are the distances to the lens and
source from the observer, respectively.  The lensing process conserves 
the surface brightness, and thus the magnification of the source star 
flux equals the ratio of the surface areas between the image and the 
unlensed source.  The magnifications of the individual images are given 
by the Jacobian of the transformation (1) evaluated at the image position, 
i.e.\
\begin{equation}
A_i = \left({1\over \vert {\rm det}\ J\vert} \right)_{\thetavec=\thetavec_i};
\qquad 
        {\rm det}\ J = \left\vert{\partial\thetavec_{\rm S}
        \over \partial\thetavec}
\right\vert.
\end{equation}
Then, the total magnification and the centroid shift vector are obtained
respectively by $A=\sum_{i=1}^{N_I} A_i$ and
$\deltavec=\sum_{i}^{N_I} A_i \thetavec_i / A
-\thetavec_{\rm S}$,
where $N_I$ is the total number of images.

When a single source is lensed by a single point-mass lens (standard 
event), the lens equation is easily solvable and yields two solutions.
The resulting images appear at the positions 
\begin{equation}
\thetavec_{\pm} = {\theta_{\rm E}\over 2} 
\left[ u \pm (u^2+4)^{1/2} \right] {\uvec\over u},
\end{equation}
and have amplifications
\begin{equation}
A_\pm = {1\over 2}
\left[ {u^2+2\over u(u^2+4)^{1/2}}\ \pm\ 1\right],
\end{equation}
where $\uvec=(\thetavec_{\rm S}-\thetavec_{\rm L})/\theta_{\rm E}$ is the
dimensionless lens-source separation vector normalized by $\theta_{\rm E}$.
The separation vector is related to the single lensing parameters by
\begin{equation}
\uvec = \left( {t-t_0\over t_{\rm E}}\right)\ \hat{\xvec}
        \ \pm\ \beta\ \hat{\yvec},
\end{equation}
where $t_{\rm E}$ is the Einstein ring radius crossing time (Einstein 
time scale), $\beta$ is the closest lens-source separation (impact 
parameter)\footnote{Note that the sign `$\pm$' in front of $\beta$ in 
eq.\ (6) is used because $\beta$ is positive definitive.}, $t_0$ is 
the time at the moment of the closest approach, and the unit vectors
$\hat{\xvec}$ and $\hat{\yvec}$ are parallel with and normal to the 
direction of the relative lens-source transverse motion, respectively. 
Then, the brighter image (major image), denoted by the subscript `$+$', 
is located outside of the Einstein ring and the other fainter image 
(minor image), denoted by the subscript `$-$', is located inside of the 
ring along the line connecting the lens and the source.  The two images 
formed by the lens cannot be resolved, but one can measure the total 
magnification and the shift of the source image centroid with respect to 
its unlensed position, which are represented respectively by

\begin{equation}
A = A_+ + A_- = {u^2+2\over u\sqrt{u^2+4}}.
\end{equation}
and
\begin{equation}
\deltavec = {A_+\thetavec_+ + A_-\thetavec_- \over A} - \uvec
= {\uvec\over u^2+2}\ \theta_{\rm E}.
\end{equation}
The light curve of a standard event has a smooth and symmetric shape, 
where its height (peak magnification) and width (duration) are determined 
respectively by $\beta$ and $t_{\rm E}$.  The centroid shift traces an 
elliptical trajectory with a semi-major axis $a=\theta_{\rm E}/2(\beta^2+2)$
and an eccentricity $\varepsilon=[(\beta^2/2)+1]^{-1/2}$ \citep{jeong99}.

\subsection{Planetary Microlensing Events}
The lens system with a planet is described by the formalism of the binary
lens system with a very low mass companion.  For this case, the lens 
equation can be represented as a fifth degree polynomial \citep{witt95} 
and numerical solution of the equation yields three or five solutions 
depending on the source position with respect to the lenses.

The main new features of binary lenses relative to single point-mass lenses
are caustics.  These are the closed curves on the source plane where a point 
source is infinitely magnified.  Hence, significant deviations both in the 
light and astrometric curves occur when the source passes the region close 
to the caustics.  The size of the caustics, and thus the planet detection 
probability, depends both on the mass ratio, $q$, and the projected separation 
(normalized by $\theta_{\rm E}$), $b$, between the planet and the primary 
lens.  The caustic size decreases with the decreasing value of $\sqrt{q}$, 
and thus planet-induced perturbations last only a short period of time.  
For a given mass ratio, the caustic size is maximized when the projected 
star-planet separation is in the range $0.6 \lesssim b\lesssim 1.6$, 
the `lensing zone'.

There are two types of planetary perturbations: those which perturb the 
major image of the source formed by the primary lens, and those which 
perturb the minor image \citep{gould92}.  Photometrically, major image 
perturbations are characterized by positive deviations from the single 
lensing light curve, while minor image perturbations are characterized 
by negative deviations \citep{gaudi98}.

\subsection{Binary-Source Events}

Unlike the non-linear behavior of binary-lens events, binary-source 
lensing is simple because the lensing behavior involved with each source 
can be treated as an independent single source event \citep{griest92}.
Then the light and astrometric curves of a binary-source event are 
represented by
\begin{equation}
A = {A_1+{\cal R} A_2 \over 1+{\cal R}}, 
\end{equation}
and
\begin{equation}
\deltavec=
{A_1(\uvec_1+\deltavec_1) + {\cal R} A_2 (\uvec_2+\deltavec_2) 
\over A_1 + {\cal R} A_2} - 
{\uvec_1 + {\cal R} \uvec_2 \over 1+ {\cal R}},
\end{equation}
where $\uvec_i$ are the separation vectors between the lens and the
individual source components, ${\cal R}$ is their (unlensed) flux ratio, 
$A_i=(u_i^2+2)/u_i(u_i^2+4)^{1/2}$ and 
$\deltavec_i=\uvec_i\theta_{\rm E}/(u_i^2+2)$ are the magnifications
and centroid shifts of the individual single source events (denoted by 
the subscripts $i=1$ and 2) \citep{han01a}.  We note that the reference 
position of the centroid shift measurements for the binary-source event 
is the center of light between the unlensed source components, i.e, the 
second term in eq.\ (10).

\begin{figure*}[t]
\plotfiddle{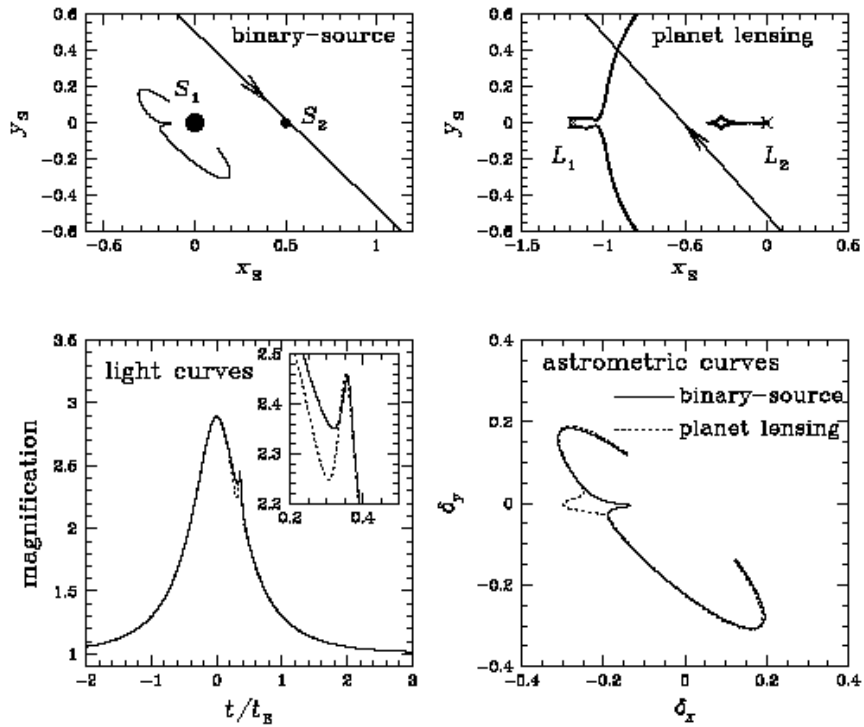}{0.0cm}{0}{65}{65}{-195}{-430}
\vskip10.5cm
\caption{
The light (lower left panel) and astrometric curves (lower right panel)
resulting from an example case of binary-source and planet microlensing
events suffering from the photometric planet/binary-source degeneracy.
The upper panels show the lens system geometries of the binary-source
(upper left panel) and the planetary lensing events (upper right panel).
The geometry of the binary-source event is represented with respect to 
the sources, denoted by $S_1$ (brighter source) and $S_2$ (fainter source), 
and the straight line with an arrow represents the lens trajectory.
On the other hand, the geometry of the planetary lensing event is 
represented with respect to lenses, denoted by $L_1$ (primary lens) 
and $L_2$ (planet), and the straight lines with an arrow represents the 
source trajectory.  The astrometric curve of the binary-source event 
is additionally presented in the upper left to show the centroid motion 
with respect to the source positions.  The curves in the upper 
right panel are the caustics (diamond shaped figure) and the critical 
curve of the planet lens system.
}
\end{figure*}

\section{Planet/Binary-source Degeneracy}
If the sources of a binary-source event have a small flux ratio and the 
fainter source passes close to the lens, the resulting light curve can 
mimic that of a planetary lensing event.  The perturbation produced by the 
binary source companion is always positive, and thus it imitates major image 
perturbations, which comprises majority of planet-induced perturbations.  
In Figure 1, we illustrate this degeneracy by presenting the light curves 
(lower left panel) resulting from an example case of binary-source (solid 
curve) and planetary lensing (dotted curve) events suffering from the 
degeneracy.  In the upper panels, we also present the lens system geometries 
of the events.  We note that while the geometry of the binary-source event 
is represented with respect to the sources, the geometry of the planetary 
lensing event is represented with respect to the lenses.  For the 
binary-source event, the flux ratio and the separation between the sources 
are ${\cal R}=5\times 10^{-3}$ and $b=0.5$, respectively.  For the planetary 
lensing event, the mass ratio and the separation between the lenses are 
$q=10^{-3}$ and $b=1.183$, respectively.  We note that the light curves 
are identical to those presented in Fig.\ 1 of \citet{gaudi98}.

To mimic the light curve of a planetary lensing event, a binary-source event 
should satisfy the specific requirements of the small flux ratio and the 
close passage of the lens to the fainter source.  \citet{gaudi98} estimated 
the probabilities to satisfy these requirements and found that for binary 
sources with separations of $0.5\lesssim b \lesssim 1.5$ the probabilities 
range from a few percent to $\sim 30$ percent for binaries with magnitude 
differences of $\Delta V\sim 4-7$.  These probabilities are of the same 
order of magnitude as the probabilities of detecting anomalies induced by 
Earth-mass to Jupiter-mass planets located in the lensing zone \citep{gould92, 
bennett96, peale01}.  Therefore, binary sources can seriously contaminate 
a sample of planet candidates detected by microlensing.

\citet{gaudi98} discussed several methods to break the degeneracy.  
These include accurate and dense photometric sampling and multi-band 
(optical/infrared) photometric measurements of color changes (or 
spectroscopic measurements of the variation in spectrum) during 
perturbations.  The first method utilizes the differences in the detailed 
lensing light curves between the two degenerate cases of events.  However,
applying this method is significantly limited because the differences are,
in general, very small and even noticeable differences last a very short 
period of time.  The other method utilizes the differences in the 
behaviors of color or spectral changes occurred during the perturbations.   
However, this method also has significant limitations because the expected 
color or spectral changes are very small.

However, if events are monitored from astrometric follow-up observations,
the planet/binary-source degeneracy can be unambiguously broken due to the 
fundamental differences between the astrometric lensing behaviors of 
binary-source and planetary lensing events.  To demonstrate this, in the 
lower right panel of Figure 1, we present the astrometric curves resulting 
from the photometrically degenerate case of binary-source (solid curve) 
and planetary lensing (dotted curve) events whose light curves are presented 
in the lower left panel.  One finds that while the planet-induced 
perturbation points away from the opening of the unperturbed astrometric 
curve (outward perturbation), the perturbation induced by the binary source 
companion points towards the opening (inward perturbation).  These trends 
of astrometric perturbations for the binary-source and planetary lensing 
events are not specified only for the presented example events, but are 
generic properties of these types of events.  From the investigation of 
of the properties of planet-induced astrometric perturbations, \citet{han01b} 
found that both cases of the planet's perturbations of major and minor 
images result in the same outward astrometric deviations.\footnote{For 
more examples of planet-induced astrometric perturbations, see Fig.\ 5 
(for major image perturbations) and Fig.\ 7 (for major image perturbations) 
of \citet{han01b}.} The characteristic inward astrometric perturbations 
of binary-source events occur due to the specific geometry of the lens 
systems.  This can be seen in the upper left panel of Fig.\ 1, where we 
additionally plot the astrometric curve of the binary-source event to 
see the source image centroid motion with respect to the source positions.  
When the lens is away from the source companion, the motion is well 
described by that of the event with the single brighter source and the 
centroid follows an elliptical trajectory.  During this time, the image 
centroid is located on the opposite side of the lens with respect to the 
unlensed position of the brighter source.  When the lens approaches the 
source companion, on the other hand, the companion is highly amplified 
and the centroid motion is perturbed.  Since the lens is close to the 
companion, this perturbation is directed towards the lens, which is 
opposite direction compared to the centroid location without the companion.  
Therefore, binary-source events producing light curves resembling those 
produced by planetary lens system {\it always} result in astrometric 
curves with inward perturbations.

\section{Conclusion}
We have shown that with additional information from astrometric lensing 
observations one can unambiguously break the photometric degeneracy between
binary-source and planetary lensing perturbations.  This is possible 
because binary-source events producing light curves imitating those of 
planetary lensing events result in astrometric perturbations pointing 
towards the opening of the unperturbed astrometric curves, while 
astrometric perturbations induced by planets point away from the opening.
Therefore, astrometric lensing observations by using future high precision 
interferometers will be important for solid confirmation of planet 
detections.

\acknowledgements
This work was supported by a grant (2001-DS0074) from the Korea 
Research Foundation (KRF).


\clearpage

\end{document}